# Microcavity-integrated graphene photodetector

Marco Furchi[1], Alexander Urich[1], Andreas Pospischil[1], Govinda Lilley[1], Karl Unterrainer[1], Hermann Detz[2], Pavel Klang[2], Aaron Maxwell Andrews[2], Werner Schrenk[2], Gottfried Strasser[2], and Thomas Mueller[1,*]

[1]Vienna University of Technology, Institute of Photonics, Gußhausstraße 27-29, 1040 Vienna, Austria

[2]Vienna University of Technology, Center for Micro- and Nanostructures, Floragasse 7, 1040 Vienna, Austria

**The monolithic integration of novel nanomaterials with mature and established technologies has considerably widened the scope and potential of nanophotonics. For example, the integration of single semiconductor quantum dots into photonic crystals [1] has enabled highly efficient single-photon sources. Recently, there has also been an increasing interest in using graphene [2] – a single atomic layer of carbon – for optoelectronic devices [3–15]. However, being an inherently weak optical absorber (only ≈2.3 % absorption), graphene has to be incorporated into a high-performance optical resonator or waveguide to increase the absorption and take full advantage of its unique optical properties. Here, we demonstrate that by monolithically integrating graphene with a Fabry-Pérot microcavity, the optical absorption is 26-fold enhanced, reaching values >60 %. We present a graphene-**

*Email: thomas.mueller@tuwien.ac.at*

**based microcavity photodetector with record responsivity of 21 mA/W. Our approach can be applied to a variety of other graphene devices, such as electro-absorption modulators, variable optical attenuators, or light emitters, and provides a new route to graphene photonics with the potential for applications in communications, security, sensing and spectroscopy.**

In principle, the light-matter interaction in graphene is strong. The optical absorption coefficient of single-layer graphene [16] is $-\ln(1-\pi\alpha)/d \approx 7\times10^{5}\,\mathrm{cm}^{-1}$, independent of wavelength ($d$ = 0.335 nm is the thickness of graphene and $\alpha$ is the fine structure constant). At the technologically important wavelengths of 850, 1300 and 1550 nm, this value is between 1 and 3 orders of magnitude higher than that of conventionally used semiconductor materials such as $In_{0.53}Ga_{0.47}As$, GaAs or Ge [17]. Nevertheless, due to the short interaction length, a layer of graphene absorbs only $\pi\alpha$ = 2.3 % of the incident light [16]. Whereas the weak optical absorption is beneficial to devices such as LCD screens [4, 5], solar cells [6–8], or organic light-emitting diodes [9], it is detrimental to active optoelectronic devices [10–15], where a strong light-matter interaction is desired. Therefore, strategies need to be developed to increase the interaction length of light with graphene and enhance the optical absorption. Recently, the integration of graphene with an optical waveguide allowed the increase of the interaction length through coupling between the evanescent waveguide mode and graphene, resulting in -3 dB (50 %) absorption in a ≈30 μm long device [13]. However, highly efficient coupling of light into a single-mode optical waveguide is challenging and in many cases surface coupling is preferred.

The graphene microcavity devices demonstrated in this letter, benefit from the large increase of the optical field inside a resonant cavity, giving rise to increased absorption. The field enhancement occurs only at the design wavelength, whereas off-resonance wavelengths are rejected by the cavity, making these devices promising for wavelength division multiplexing (WDM) systems [18]. Cavity enhanced devices have a long history in III-V optoelectronics [19–21]. However, monolithic integration of carbon nano-materials with optical cavities is challenging and reports are rare [22].

A graphene microcavity photodetector (GMPD) is shown schematically in Figure 1(a). As nominal operating wavelength we have chosen $\lambda_c$ = 850 nm, a wavelength that is often used in low-cost multi-mode fiber data links [23]. However, due to the broad absorption range of graphene [16, 24], this concept can be extended to any wavelength from the mid-infrared to the ultraviolet, provided that a low-loss optical cavity can be realized at the respective wavelength [25, 26]. In our device, two distributed Bragg mirrors, consisting of quarter-wavelength thick layers of alternating materials with varying refractive indices, form a high-finesse planar cavity. Bragg mirrors are ideal choices for microcavity optoelectronic devices because, unlike with metal mirrors, the reflectivity can be very well controlled and can reach values near unity. The Bragg mirrors are made of large band gap materials that are non-absorbing at the detection wavelength $\lambda_c$. The absorbing graphene layer is sandwiched between these mirrors. A $Si_3N_4$ buffer layer ensures that the maximum of the field amplitude occurs right at the position where the graphene sheet is placed. The bottom mirror is formed from multiple periods (25 pairs) of alternating AlAs and $Al_{0.10}Ga_{0.90}As$ layers (GaAs would be absorbing at 850 nm) with thicknesses of 70 nm and 61 nm, respectively. It is grown by molecular beam epitaxy (MBE) on a GaAs substrate. The refractive index contrast of AlAs and $Al_{0.10}Ga_{0.90}As$ gives a mirror with reflectivity $R_{bottom}$ > 99 % in a broad spectral range around 850 nm [see inset of Figure 1(c)]. The top mirror is made of 7 pairs of $SiO_2$ and $Si_3N_4$ layers with thicknesses of 147 nm and 113 nm, respectively. Its nominal reflectivity is 89 %. The graphene is electrically contacted by lithographically defined Ti/Au electrodes. Weak doping of the substrate and the bottom Bragg mirror allows electrostatic gating of the graphene channel and measurements of the electrical characteristics. The devices exhibit the typical V-shaped conductance versus gate bias with weak unintended doping (<10 V Dirac point shift) and negligible hysteresis (<2.5 V). However, we observe a strong mobility reduction to typically a few hundred $cm^2/Vs$, as compared to the ≈5000 $cm^2/Vs$ that are obtained in "conventional" graphene devices [10, 11]. Similar impact of the dielectric environment on the mobility was previously reported for high-frequency graphene transistors [27]. More information on device fabrication is presented in the Methods section. The detailed device structure is provided as Supplementary information.

The device design was optimized using the transfer matrix method. In our simulation, graphene is described by a complex refractive index $n(\lambda) = 3.0 + i\frac{C_1}{3}\lambda$, where $C_1$ = 5.446 $\mu m^{-1}$ and $\lambda$ is the wavelength [28]. The other materials are modeled as loss-less dielectric materials with refractive indices reported in the Supplementary Information. Figure 1(b) shows the electric field distribution in the device for normal incidence light at the design wavelength of $\lambda_c$ = 850 nm. The standing wave pattern arises from interference of the counter-propagating incident and reflected waves. The arrow indicates the spatial position of the graphene. It is obvious that the origin of the absorption enhancement is the ≈6.5-fold increased electric field amplitude inside the cavity, which causes more energy to be absorbed. An equivalent interpretation is that the photons bounce between the bottom and top mirrors and thus passes multiple times through the graphene sheet as illustrated in Figure 1(a). We calculate the wavelength-dependent absorption $A(\lambda)$ according to $A(\lambda) = 1 - R(\lambda) - T(\lambda)$, where $R(\lambda)$ is the (intensity) reflection and $T(\lambda)$ denotes the transmission. The model allows us to optimize the reflectivity $R_{top}$ of the top mirror. As shown in Figure 1(c), the absorption increases with increasing reflectivity, reaches a maximum of 98 % for 7 $SiO_2/Si_3N_4$ layer pairs (corresponding reflectivity $R_{top}$ = 89 %) and drops to zero as $R_{top}$ approaches 100 %. This behavior can also be understood intuitively. For small $R_{top}$, the cavity is too lossy and the field enhancement is small. For $R_{top}$ = 100 %, on the other hand, all the light is reflected on the surface and cannot enter into the cavity. For bi-layer graphene we find an optimum of 6 instead of 7 $SiO_2/Si_3N_4$ layer pairs.

The performance of GMPDs depends critically on the optical quality of the mirror and buffer layer materials, which must be non-absorbing at the detection wavelength $\lambda_c$. We have therefore measured the wavelength dependent optical reflection of a millimeter-sized spot on the sample. The corresponding spectrum (shown in Figure 2) is a typical reflectivity spectrum of a Bragg mirror stack, but exhibits an extra dip at $\lambda$ = 850 nm in the stop band. The dip originates from absorption of the Fabry-Pérot cavity mode. Its depth is less than 6 % and supposedly stems mainly from absorption in the (multi-layer) graphene flakes that are randomly distributed over the sample (a few % surface coverage). The weak absorption is an evidence of the high optoelectronic material quality.

Let us now turn to the photocurrent measurements. Figure 3(b) shows a microscope image of a single-layer GMPD together with the measurement circuit. The graphene channels of our devices are typically 5 μm long and several microns wide. A bias voltage $V_{\mathrm{Bias}}$ = 2 V is applied to one of the leads. The other lead is connected to a transimpedance low-noise pre-amp whose output signal is fed into a lock-in amplifier. The gate electrode (substrate) remains unbiased. The output of a tunable continuous-wave Ti:sapphire laser is set to 850 nm wavelength and is focused with an objective lens to a ≈2 μm diameter spot on the sample. The optical power was kept low enough ($P$ = 50 μW) to avoid heating of the sample and reduce the influence of thermo-electric effects [29–31]. A photocurrent map [shown in Figure 3(a)] is recorded by scanning the laser beam across the sample. The incident light is modulated at 400 Hz using a mechanical chopper. This technique has previously been used to study the potential profiles in graphene transistors, where photocurrents of opposite sign at the metal/graphene interfaces were observed [32–35]. Due to the different biasing condition, the band bending [see Figure 3(a)] in our experiment is determined by the externally applied voltage, rather than by the metal/graphene contacts. We therefore observe only a single photocurrent peak approximately in the center of the device, whose polarity is determined by the sign of $V_{\mathrm{B}}$.

Although this biasing condition does not allow zero dark current operation, it reduces the influence of the metal electrodes on the shape of the photocurrent spectrum.

In Figure 3(c) we present the spectral response of the device. The dashed lines are results of the transfer matrix calculation; the solid lines are measurement results. The reflectivity spectrum is measured by focusing the Ti:sapphire laser output to a small spot in the center of the graphene sheet and by tuning the laser wavelength between 830 and 900 nm. The result is shown in Figure 3(c) as solid red line. At the cavity resonance (855 nm), more than 60 % of the light is absorbed in the graphene – a 26-fold absorption enhancement as compared to the 2.3 % absorption of free-standing graphene [16]. The slight deviation from design wavelength ($\lambda_c$ = 850 nm) is caused by small non-uniformities (≈0.6 %) in optical layer thicknesses of the buffer and top mirror layers. We accounted for this deviation in our simulation. At $\lambda$ = 888 nm another reflection dip is observed, which is, however, not related to absorption in the graphene but stems from larger transmission outside the stop band of the Bragg mirror stack (see green line). The measurement data are well reproduced by the simulation (dashed red line) if we consider spectral broadening due to the finite numerical aperture (*NA* = 0.28) of the objective lens by numerically averaging over all incidence angles between 0 and $\vartheta_{max}$ = Arcsin(*NA*) ≈ 16°. The solid blue line in Figure 3(c) shows the spectral photocurrent response of the device. It peaks at $\lambda$ = 855 nm wavelength and exhibits a spectral width of $\Delta\lambda$ = 9 nm (full width at half maximum – FWHM). Its shape follows closely the calculated absorption (dashed blue line), demonstrating that the absorbed light is efficiently converted into photocurrent. From the quality factor of the cavity, $Q = \lambda / \Delta\lambda = 95$, we obtain a photon lifetime of $\tau = Q\lambda/(2\pi c)$ = 43 fs, only. The microcavity does hence not affect the potentially high bandwidth [11] of graphene photodetectors. The inset shows the calculation results for $NA$ = 0 ($\vartheta_{max}$ = 0°), i.e. normal incidence light. In this case, the absorption would be as large as 98 %.

In Figure 4 we show the results obtained from a bi-layer graphene device. The meaning of the curves is the same as in Figure 3(c). Again, we observe a strong photoresponse at the cavity resonance (864.5 nm, in this case). A peak photocurrent of $I$ = 1.05 μA is obtained at $P$ = 50 μW excitation power, which translates into a photoresponsivity of $S = I/P$ = 21 mA/W. To our knowledge, this is the highest value ever reported for a graphene photodetector. Also shown in Figure 4 (solid red line) is the response of a "conventional" bi-layer graphene photodetector, i.e. a device without cavity. It consists of bi-layer graphene on a Si wafer with 300 nm thick $SiO_2$ and Ti/Au electrodes. For a fair comparison, the geometrical dimensions (particularly the channel length) of the device are similar to those of our GMPD devices and also the biasing conditions are the same. The response of the conventional device is approximately independent of wavelength, but more than an order of magnitude weaker than that of the microcavity enhanced device.

In conclusion, we have demonstrated that the responsivity of a graphene photodetector can be increased, by integrating the graphene sheet in a high-finesse planar optical cavity. A record responsivity of $S$ = 21 mA/W is achieved. The devices show a photoresponse only at the design wavelength, making them promising for wavelength division multiplexing. The concept of enhancing the light-matter interaction in graphene by use of an optical microcavity is not limited to photodetectors alone. It can be applied to a variety of other devices such as electro-absorption modulators, variable optical attenuators, and possibly future light emitters. Our demonstration also shows that graphene can be monolithically integrated with other, more established materials and technologies to form novel, highly complex devices.

## Methods

Sample fabrication began with an n-doped GaAs wafer. On top of the wafer, a 300 nm thick GaAs smoothing layer, followed by 25 pairs of AlAs/$Al_{0.10}Ga_{0.90}As$ layers (70/61 nm; all weakly n-doped) were grown by molecular beam epitaxy (MBE). We then deposited a 111 nm thick $Si_3N_4$ layer by plasma enhanced chemical vapor deposition (PECVD) using $SiH_4$ and $NH_3$ precursor gases at a substrate temperature of 300 °C. Graphene flakes were deposited by mechanical exfoliation. Single- and bi-layer flakes were visually located with a microscope and subsequently confirmed to be single- or bi-layers with Raman spectroscopy. Source and drain Ti/Au (10/20 nm) electrodes were then deposited by laser lithography, electron-beam evaporation of the metals, and lift-off. In a second lithography step, 70 nm thick Au contact pads were patterned. Overnight thermal annealing at 150 °C under vacuum was performed to remove unintentional doping, including water molecules. The annealing was performed in the PECVD chamber, so that the subsequent $SiO_2$/$Si_3N_4$ (147/113 nm; 7 pairs) layers could be deposited without bringing the sample back to atmosphere. The PECVD deposition was again performed at a substrate temperature of 300 °C. $SiH_4$ and $N_2O$ were used as precursor gases for the $SiO_2$ deposition. For $Si_3N_4$ deposition, the same process parameters as above were used. During all PECVD deposition processes, dummy Si wafers were placed along with the sample into the chamber. This allowed precise determination of the film thicknesses by optical measurements (thin film thickness measurement system; ellipsometry). Finally, using reactive ion etching (RIE), contact windows were etched in the $SiO_2/Si_3N_4$ top mirror at the position of the contact pads.

## Acknowledgements

This work was supported by the Austrian Science Fund (FWF).

## Figure captions

***Figure 1.*** **Graphene microcavity photodetector.** (a) Schematic drawing of the device. Distributed Bragg mirrors form a high-finesse optical cavity. The incident light is trapped in the cavity and passes multiple times through the graphene. The graphene sheet is shown in red, the metal contacts in yellow. (b) Electric field amplitude inside the cavity. (c) Calculated dependence of optical absorption in a single-layer graphene sheet on the reflectivity of the top mirror. The numbers next to the symbols indicate the number of $SiO_2/Si_3N_4$ layer pairs that are necessary to achieve the respective reflectivity. Inset: Measured reflectivity of the $AlAs/Al_{0.10}Ga_{0.90}As$ bottom mirror.

***Figure 2.*** **Microcavity resonance.** Reflectivity of the sample. The dip at 850 nm wavelength originates from absorption of the Fabry-Pérot cavity mode.

***Figure 3.*** **Spatially and spectrally resolved photocurrent response.** (a) Photocurrent map taken at a bias voltage of $V_{Bias}$ = 2 V between the source and drain electrodes. The gate electrode (substrate) remains unbiased. The dashed lines indicate the source and drain electrodes. The schematic above the photocurrent map illustrates the band diagram under this biasing condition. (b) Microscope image of a graphene photodetector and electrical setup for photocurrent measurements. The scale bar is 5 μm long. (c) Spectral response of the single-layer graphene device. The dashed lines show calculation results: reflection $R$ (red), transmission $T$ (green), and absorption $A$ (blue). The solid lines are measurement results: reflection (red); photocurrent (blue). A strong and spectrally narrow photoresponse is observed at the cavity resonance (855 nm wavelength). Inset: Theoretical result for normal incidence light.

***Figure 4.*** **Comparison between microcavity and conventional photodetector.** The meaning of the curves is the same as in Figure 3(c), but the results are shown for a bi-layer graphene device. A maximum responsivity of 21 mA/W is achieved. In addition, the spectral photoresponse of a conventional (without cavity) bi-layer graphene detector is shown as solid red line. The response of the conventional device is approximately independent of wavelength, but more than an order of magnitude weaker than that of the microcavity device.

## References

1. Badolato, A. *et al.* Deterministic Coupling of Single Quantum Dots to Single Nanocavity Modes. *Science* 308, 1158–1161 (2005).
2. Novoselov, K.S. *et al.* Electric Field Effect in Atomically Thin Carbon Films. *Science* 306, 666–669 (2004).
3. Bonaccorso, F., Sun, Z., Hasan, T., Ferrari, A. C. Graphene photonics and optoelectronics. *Nat. Photon.* 4, 611–622 (2010).
4. Blake, P. *et al.* Graphene-Based Liquid Crystal Device. *Nano Lett.* 8, 1704–1708 (2008).
5. Bae, S. *et al.* Roll-to-roll production of 30-inch graphene films for transparent electrodes. *Nat. Nanotech.* 5, 574–578 (2010).
6. Wang, X. *et al.* Transparent carbon films as electrodes in organic solar cells. *Angew. Chem.* 47, 2990–2992 (2008).
7. Wu, J. *et al.* Organic solar cells with solution-processed graphene transparent electrodes. *Appl. Phys. Lett.* 92, 263302 (2008).
8. De Arco, L.G. et al. Continuous, Highly Flexible, and Transparent Graphene Films by Chemical Vapor Deposition for Organic Photovoltaics. *ACS Nano* 4, 2865–2873 (2010).
9. Wu, J. *et al.* Organic Light-Emitting Diodes on Solution-Processed Graphene Transparent Electrodes. *ACS Nano* 4, 43–48 (2010).
10. Mueller, T., Xia, F. & Avouris, Ph. Graphene photodetectors for high-speed optical communications. *Nat. Photon.* 4, 297–301 (2010).
11. Xia, F., Mueller, T., Lin, Y., Valdes-Garcia, A. & Avouris, Ph. Ultrafast graphene photodetector. *Nat. Nanotech.* 4, 839–843 (2009).
12. Echtermeyer, T.J. *et al.* Strong plasmonic enhancement of photovoltage in graphene. *Nat. Commun.* 2, 458 (2011).
13. Liu, M. *et al.* A graphene-based broadband optical modulator. *Nature* 474, 64–67 (2011).
14. Zhang, H., Tang, D.Y., Zhao, L.M., Bao, Q.L. & Loh, K.P. Large energy mode locking of an erbium-doped fiber laser with atomic layer graphene. *Opt. Express* 17, 17630–17635 (2009).
15. Sun, Z. *et al.* Graphene Mode-Locked Ultrafast Laser. *ACS Nano* 4, 803–810 (2010).
16. Nair, R.R. *et al.* Fine Structure Constant Defines Visual Transparency of Graphene. *Science* 320, 1308 (2008).
17. Palik, E.D. (Ed.) Handbook of Optical Constants of Solids. Academic Press (1985).
18. Ishio, H., Minowa, J. & Nosu, K. Review and status of wavelength-division-multiplexing technology and its application. *J. Lightwave Technol.* 2, 448–463 (1984).

19. Ünlü, M.S. & Strite, S. Resonant cavity enhanced photonic devices. *J. Appl. Phys.* 78, 607–638 (1995).
20. Schubert, E.F., Wang, Y.-H., Cho, A.Y., Tu, L.-W. & Zydzik, G.J. Resonant cavity light-emitting diode. *Appl. Phys. Lett.* 60, 921–923 (1992).
21. Maier, T., Strasser, G. & Gornik, E. Monolithic Integration of Vertical-Cavity Laser Diodes and Resonant Photodetectors with Hybrid $Si_3N_4$–$SiO_2$ Top Bragg Mirrors. *IEEE Photon. Tech. Lett.* 12, 119–121 (2000).
22. Xia, F., Steiner, M., Lin, Y. & Avouris, Ph. A microcavity-controlled, current-driven, on-chip nanotube emitter at infrared wavelengths. *Nat. Nanotech.* 3, 609–613 (2008).
23. Pepeljugoski, P. *et al.* 15.6 Gb/s Transmission Over 1 km of Next Generation Multimode Fiber. *IEEE Phot. Technol. Lett.* 14, 717–719 (2002).
24. Mak, K.F. *et al.* Measurement of the Optical Conductivity of Graphene. *Phys. Rev. Lett.* 101, 196405 (2008).
25. Heiss, W. *et al.* Epitaxial Bragg mirrors for the mid-infrared and their applications. *Prog. Quant. Electron.* 25, 193–228 (2001).
26. Dorsaz, J., Carlin, J.-F., Gradecak, S. & Ilegems, M. Progress in AlInN–GaN Bragg reflectors: Application to a microcavity light emitting diode. *J. Appl. Phys.* 97, 084505 (2005).
27. Lin, Y. *et al.* Operation of Graphene Transistors at Gigahertz Frequencies. *Nano Lett.* 9, 422–426 (2009).
28. Bruna, M. & Borini, S. Optical constants of graphene layers in the visible range. *Appl. Phys. Lett.* 94, 031901 (2009).
29. Xu, X. *et al.* Photo-Thermoelectric Effect at a Graphene Interface Junction. *Nano Lett.* 10, 562–566 (2010).
30. Lemme, M.C. *et al.* Gate-Activated Photoresponse in a Graphene p–n Junction. *Nano Lett.* 11, 4134–4137 (2010).
31. Gabor, N.M. *et al.* Hot Carrier-Assisted Intrinsic Photoresponse in Graphene. *Science* 334, 648–652 (2011).
32. Lee, E. J. H., Balasubramanian, K., Weitz, R. T., Burghard, M. & Kern, K. Contact and edge effects in graphene devices. *Nature Nanotech.* 3, 486–490 (2008).
33. Xia, F. *et al.* Photocurrent Imaging and Efficient Photon Detection in a Graphene Transistor. *Nano Lett.* 9, 1039–1044 (2009).
34. Mueller, T., Xia, F., Freitag, M., Tsang, J. & Avouris, Ph. Role of contacts in graphene transistors: A scanning photocurrent study. *Phys. Rev. B* 79, 245430 (2009).
35. Park, J., Ahn, Y. H. & Ruiz-Vargasv, C. Imaging of Photocurrent Generation and Collection in Single-Layer Graphene. *Nano Lett.* 9, 1742–1746 (2009).

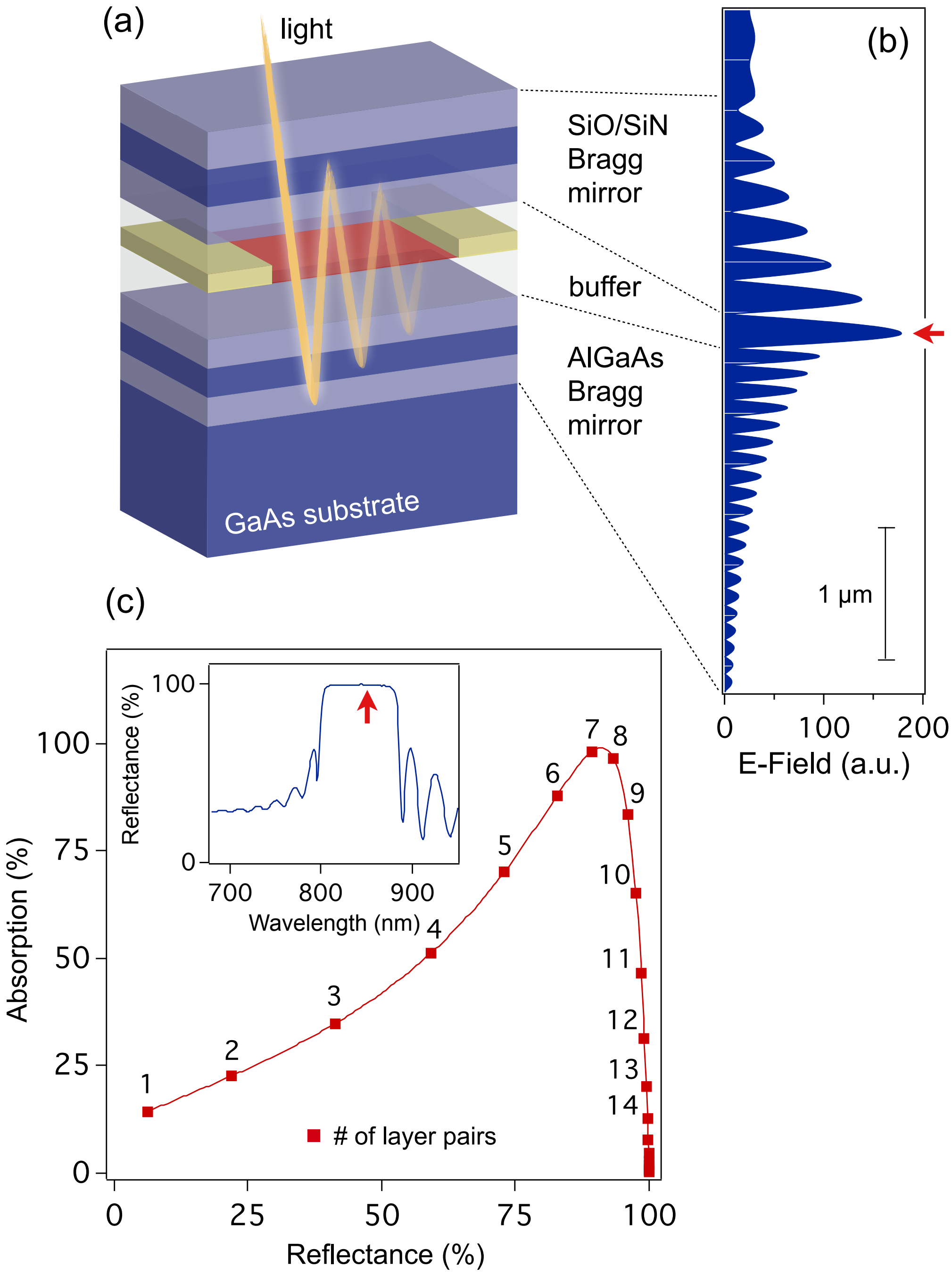

**Furchi *et al.* / Figure 1**

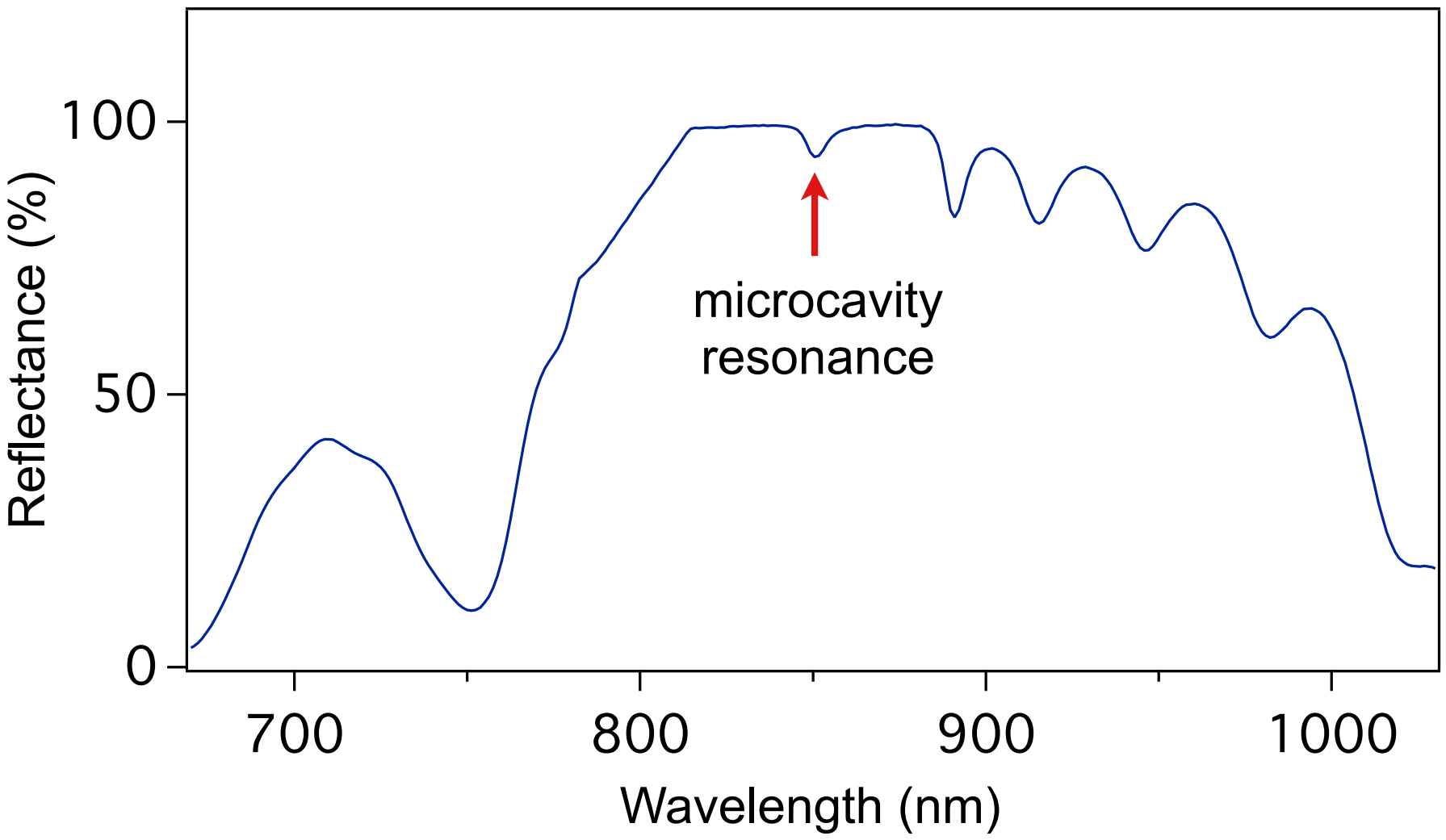

**Furchi *et al.* / Figure 2**

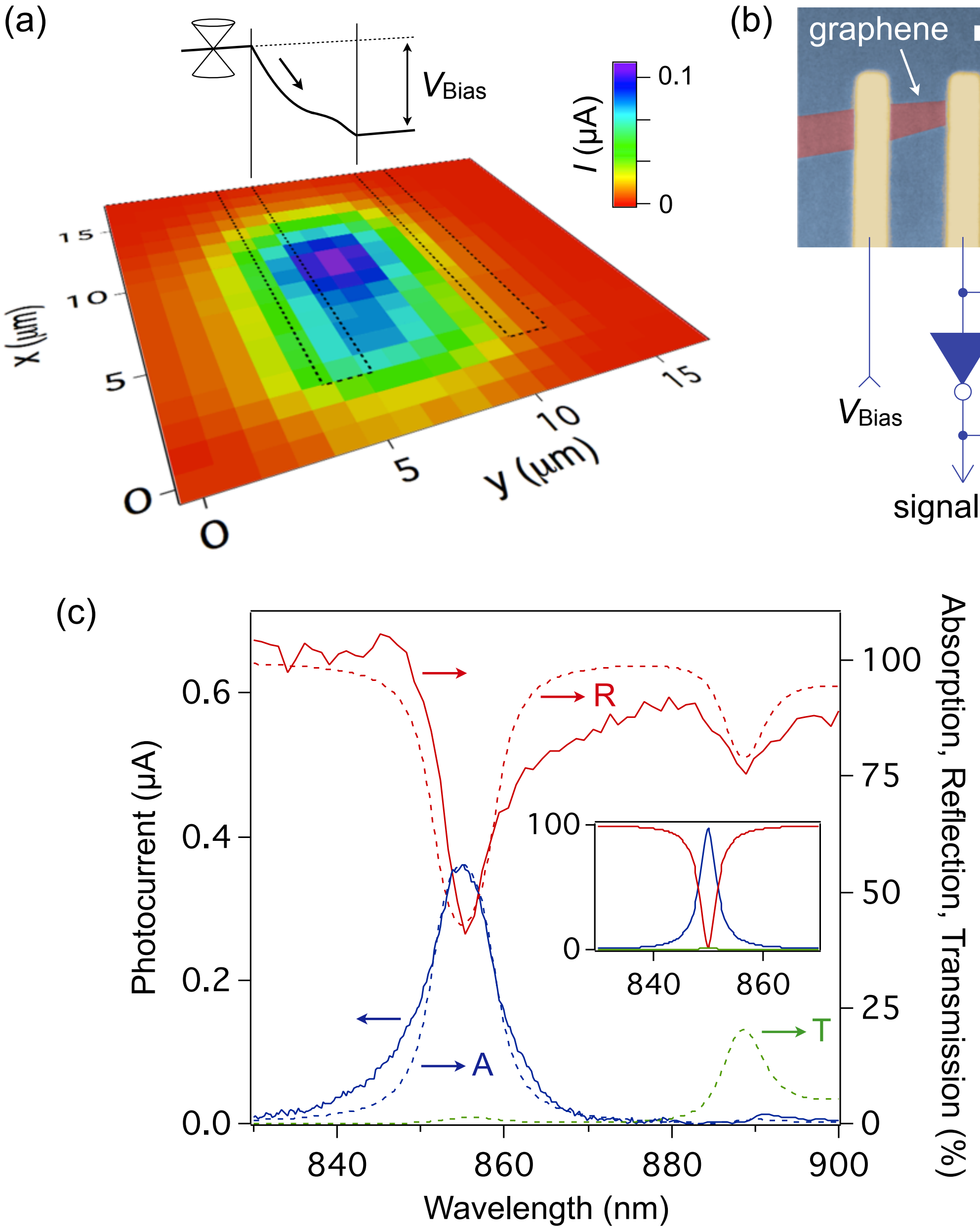

**Furchi *et al.* / Figure 3**

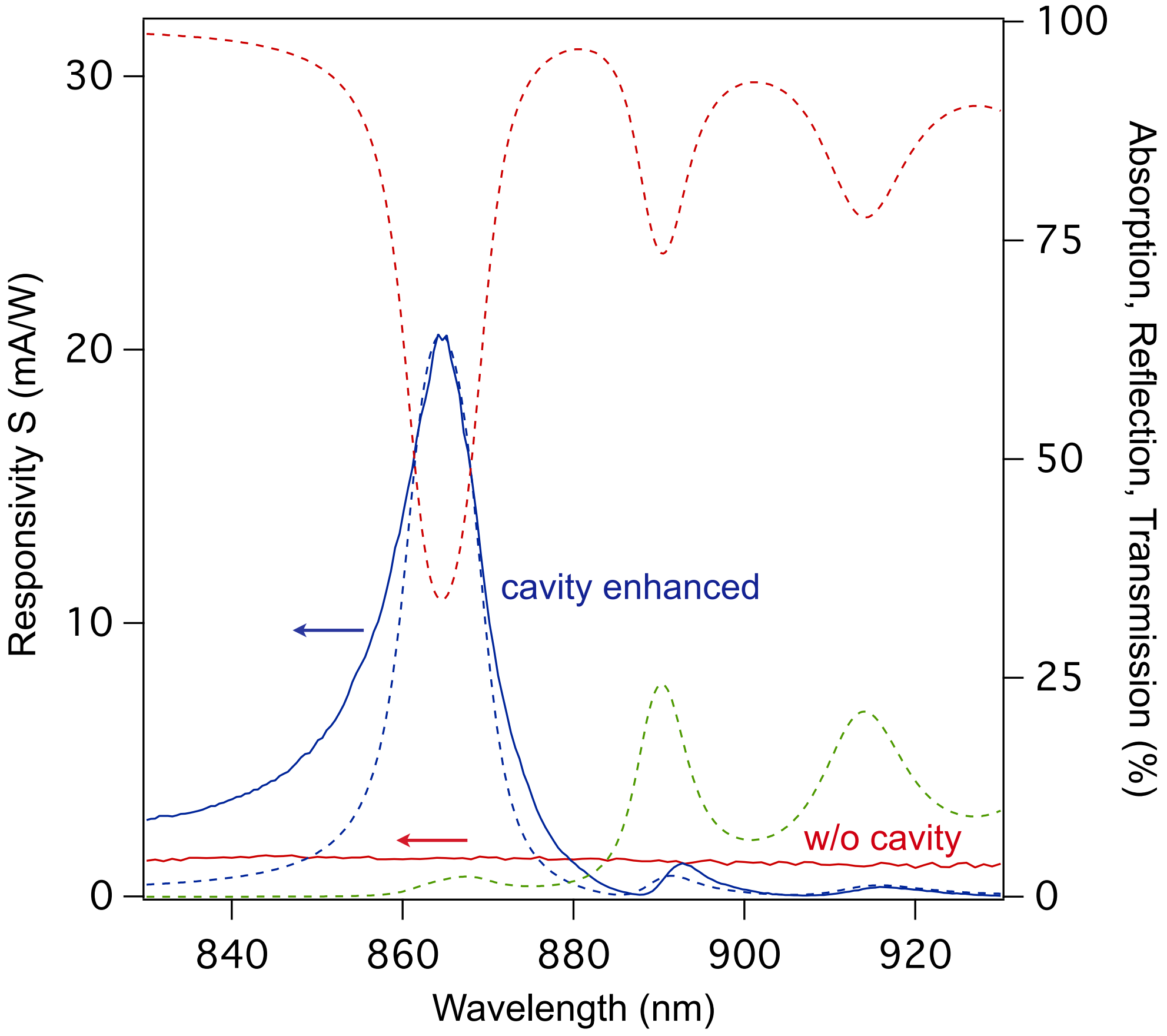

**Furchi *et al.* / Figure 4**

# Supplementary Information

| Material | Thickness (nm) | Refractive index | Band gap (eV) | Doping ($cm^{-3}$) | Comment |
|---|---|---|---|---|---|
| $Si_3N_4$ | 113 | 1.87 | 5.1 | - | top mirror *7 layer pairs* |
| $SiO_2$ | 147 | 1.45 | 8.9 | | |
| ⋮ | ⋮ | ⋮ | ⋮ | | |
| $Si_3N_4$ | 113 | 1.87 | 5.1 | | |
| $SiO_2$ | 147 | 1.45 | 8.9 | | |
| Graphene | 0.335 | $3.0+i\frac{C_1}{3}\lambda$ | 0 | - | absorbing layer |
| $Si_3N_4$ | 111 | 1.87 | 5.1 | - | buffer layer |
| $Al_{0.10}Ga_{0.90}As$ | 61 | 3.5 | 1.55 | $10^{14}$–$10^{15}$ (n) | bottom mirror *25 layer pairs* |
| AlAs | 70 | 3.05 | 2.16 | | |
| ⋮ | ⋮ | ⋮ | ⋮ | | |
| $Al_{0.10}Ga_{0.90}As$ | 61 | 3.5 | 1.55 | | |
| AlAs | 70 | 3.05 | 2.16 | | |
| GaAs | 300 | 3.55 | 1.42 | $10^{14}$ (n) | smoothing layer |
| GaAs | 625 μm | 3.55 | 1.42 | $2\times10^{18}$ (n) | substrate |

*Table S1:* Detailed device structure. $C_1 = 5.446\ \mu m^{-1}$; $\lambda$ is the wavelength.